\documentclass[9pt,conference]{IEEEtran}

\usepackage[preprint]{waspaa25}

\usepackage{bm} 

\usepackage{tabularx} 
\usepackage{multirow}  
\usepackage[table]{xcolor}

\definecolor{myblue}{HTML}{DAE8FC}
\colorlet{lightblue}{myblue!75!white}

\title{FlexSED: Towards Open-Vocabulary Sound Event Detection}

\name{
Jiarui Hai, 
Helin Wang, 
Weizhe Guo,
Mounya Elhilali
}
\address{Department of Electrical and Computer Engineering\\ Johns Hopkins University, Maryland, USA}
\begin{document}

\maketitle

\begin{abstract}
Despite recent progress in large-scale sound event detection (SED) systems capable of handling hundreds of sound classes, existing multi-class classification frameworks remain fundamentally limited. They cannot process free-text sound queries, which enable more flexible and user-friendly interaction, and they lack zero-shot capabilities and offer poor few-shot adaptability. 
Although text-query-based separation methods have been explored, they primarily focus on source separation and are ill-suited for SED tasks that require precise temporal localization and efficient detection across large and diverse sound vocabularies. In this paper, we propose FlexSED, an open-vocabulary sound event detection system. FlexSED builds on a pretrained audio SSL model and the CLAP text encoder, introducing an encoder-decoder composition and an adaptive fusion strategy to enable effective continuous training from pretrained weights. To ensure robust supervision, it also employs large language models (LLMs) to assist in event query selection during training, addressing challenges related to missing labels. As a result, FlexSED achieves superior performance compared to vanilla SED models on AudioSet-Strong, while demonstrating strong zero-shot and few-shot capabilities. We release the code and pretrained models to support future research and applications based on FlexSED.

\end{abstract}

\section{Introduction}
Sound event detection (SED) focuses on identifying and temporally locating meaningful sounds in audio recordings. Given an input audio stream, SED systems aim to determine what happened and when, recognizing events such as speech, alarms, animal sounds, or environmental noises and marking their precise start and end times \cite{mesaros2021sound, turpault2019sound}. This technology underpins a wide range of real-world applications, including smart home monitoring, health and safety surveillance, wildlife tracking, multimedia content analysis, and assistive technologies for the hearing impaired.

Advances in deep neural networks have enabled modern SED models to achieve strong performance, though mainly on small-scale corpora focused on domestic environments~\cite{turpault2019sound, nam2022frequency, shao2024fine}. More recently, breakthroughs in audio spectrogram transformers, pretrained on large-scale unlabeled data via self-supervised learning (SSL), have significantly expanded SED capabilities, allowing models to address complex and diverse acoustic scenes involving over 400 sound classes~\cite{li2024self, schmid2025effective, hershey2021benefit}. However, these approaches still operate under a closed-vocabulary assumption, treating sound event detection as a multi-class classification task constrained to a predefined set of classes established during training. This limitation not only requires users to manually specify and map target classes for each use case, complicating real-world deployment, but also restricts the model’s ability to capture rich semantic and acoustic relationships among sound categories. In addition, traditional models do not support direct inference beyond the predefined label space without retraining and incorporating additional labeled data. Moreover, adapting to new events is challenging, as it requires modifying the classification head.

While these challenges continue to constrain the flexibility and scalability of current SED systems, related advances in computer vision offer a compelling contrast. Large-scale open-vocabulary detection (OVD) has advanced rapidly in computer vision \cite{zareian2021open, minderer2023scaling}. By enabling automatic detection of arbitrary objects without predefined categories or extensive human annotation, OVD has driven image understanding toward more flexible, scalable, and adaptable recognition. These capabilities have unlocked a range of versatile applications, including evaluating text-to-image generation \cite{ghosh2023geneval} and labeling data for tasks such as visual question answering \cite{li2024convincing}. However, such large-scale open-vocabulary functionality remains largely absent in current SED systems, limiting their generalization and hindering progress in broader audio tasks, including evaluating text-to-audio generation~\cite{hai2024ezaudio, hung2024tangoflux}, enhancing audio language models ~\cite{chu2024qwen2, ghosh2024gama}, and serving as a reward signal for methods such as Direct Preference Optimization \cite{rafailov2023direct}.

Although recent efforts have explored text-prompt-based separation models~\cite{wang2025soloaudio, liu2024separate}, these approaches primarily aim to extract sound events rather than directly predict their temporal locations. Moreover, they often rely on computationally intensive methods, such as diffusion models, which are designed to improve separation quality when the target sound source is specified. This focus makes them less suitable for scenarios requiring detection across a large number of potential classes, as in SED. While text-query-based separation models have been applied to text-query-based SED~\cite{yin2025exploring}, this approach was developed on a small-scale dataset, lacks demonstrated zero-shot and few-shot capabilities, does not offer true open-vocabulary support, and incurs substantial computational overhead due to its reliance on separation processes. As a result, it remains ill-suited for more challenging scenarios involving diverse and numerous sound classes.

Motivated by recent advances in large-scale OVD and the need to overcome the limitations of both conventional multi-class classification frameworks and separation-based text-query SED, we investigate the concept of \textbf{open-vocabulary sound event detection (OV-SED)} in this work. However, pursuing OV-SED introduces several key challenges. Unlike OVD in vision, where large-scale datasets are available, the largest temporally labeled dataset for SED, AudioSet-Strong~\cite{hershey2021benefit}, contains only about 100k audio clips, limiting the amount of training data. Moreover, negative queries play a crucial role in OV-SED training to improve classification accuracy. Yet, the presence of missing labels in AudioSet-Strong complicates the reliable construction of negatives. Poorly selected negatives risk introducing noise or incorrect supervision, which can mislead the model during training and ultimately degrade performance.

\begin{figure*}[t]
  \centering
  \includegraphics[width=0.90\textwidth]{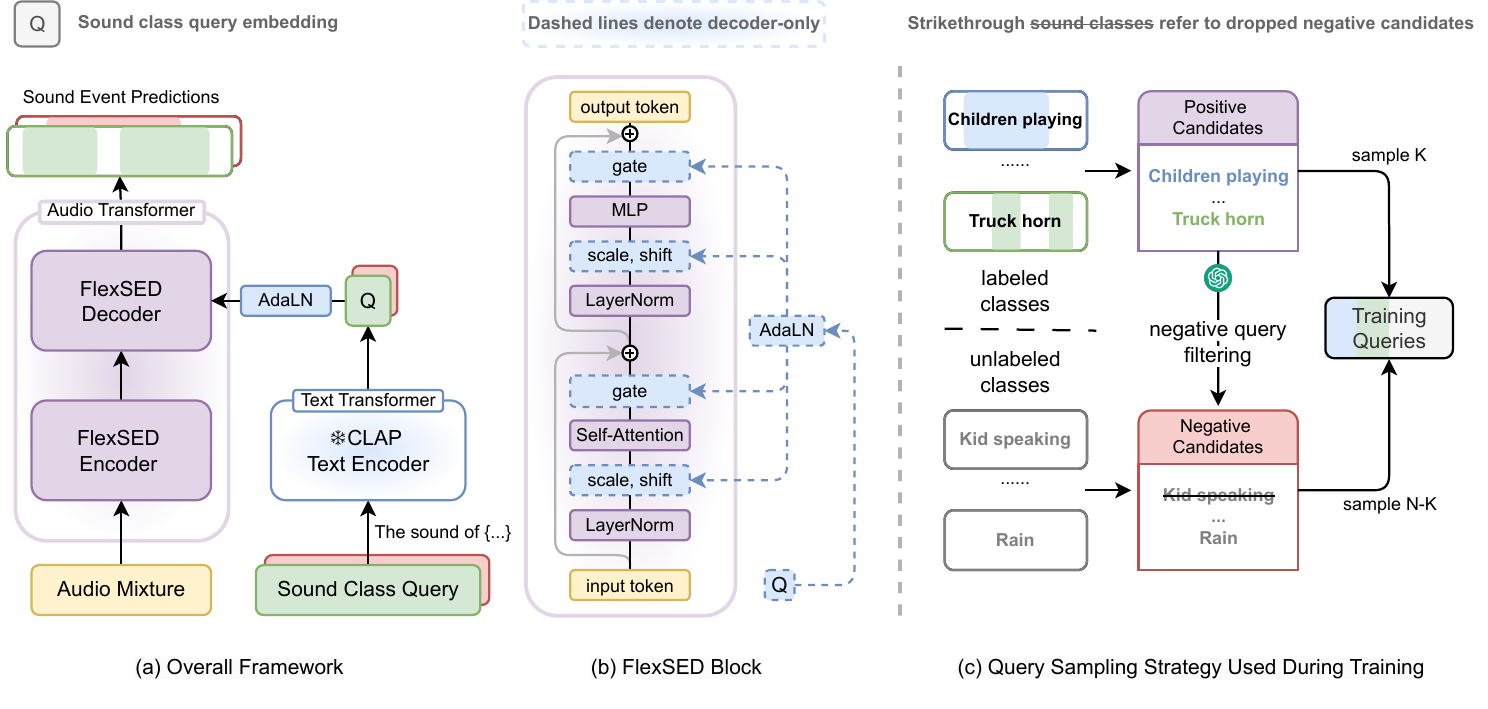}
  \caption{The framework of FlexSED.}
  \label{fig:framework}
\end{figure*}

To address these challenges and advance OV-SED, we propose FlexSED, a framework specifically designed to tackle the core issues outlined above. Its key design components are as follows:

\textbf{(1) Architecture design for prompt-aware audio modeling.} To address data scarcity and enable efficient prompt-based audio modeling, we leverage two pretrained models: a frame-based audio SSL model~\cite{dinkel2024scaling} for fine-grained acoustic representation and the CLAP text encoder~\cite{wu2023large} for sound class prompt comprehension. We decompose the audio SSL model into an audio-only encoder and an audio-text fusion decoder specialized for prompt-audio integration, balancing performance and inference efficiency. Furthermore, we introduce a seamless adaptive fusion strategy to enable continued training from pretrained weights.

\textbf{(2) Mitigating missing label issues.} To mitigate the risk of incorrect negative queries arising from incomplete labels, we propose a negative candidate filtering strategy that leverages large language models (LLMs)~\cite{achiam2023gpt} to infer semantic relationships among sound classes. This approach helps avoid conflicts between positive and negative queries, ensuring more reliable and consistent supervision.

\textbf{(3) Effectiveness and flexibility.} FlexSED surpasses conventional multi-class classification methods on the AudioSet-Strong evaluation set, while also demonstrating strong zero-shot capabilities and few-shot expandability.

\textbf{(4) Enabling practical applications.} We release\footnote{Source Code: 
 \url{https://github.com/JHU-LCAP/FlexSED}} the code, model checkpoints, and user-friendly inference tools to support a wide range of OV-SED applications, including text-to-audio evaluation, audio question answering data annotation, and more.

\section{Proposed Method}

\subsection{Overall Framework}

As shown in \hyperref[fig:framework]{\cref{fig:framework}~(a)}, OV-SED takes an audio clip along with a list of sound event candidate prompts as input. The audio signal is first processed by the encoder independently, without interacting with the candidate prompts. Meanwhile, the text prompts are processed by a text encoder. The decoder then fuses the audio-text embedding pairs in parallel, enabling interactions between candidate prompts, and produces the corresponding posterior probabilities.

\subsection{FlexSED Model Architecture}


To build a simple yet effective open-vocabulary sound event detector capable of identifying and localizing sound events from arbitrary text inputs, we propose a unified model that integrates audio and text representations. At its core, the model adopts a vanilla transformer-based SED architecture, modified to predict only the posterior corresponding to the input text prompt, rather than conventional multi-class outputs. This enables flexible, prompt-based sound event detection.

For text representation, we leverage CLAP \cite{wu2023large}, which produces rich and semantically aligned embeddings that bridge the gap between textual descriptions and audio content. For audio processing, we adopt a frame-based audio spectrogram transformer \cite{dinkel2024scaling} built on self-supervised learning, which provides strong audio representations, accelerates convergence on downstream audio and speech tasks, and operates at a latent sample rate of 25 Hz, making it well-suited for sound event detection.

However, the audio transformer does not natively support text fusion. To address this, inspired by AdaLN-Zero \cite{peebles2023scalable}, commonly used in diffusion transformers for effective feature fusion, we introduce AdaLN into the pre-trained audio SSL blocks, as illustrated in \hyperref[fig:framework]{\cref{fig:framework}~(b)}. Unlike AdaLN-Zero, which initializes the residual scaling gate to zero, we initialize this gate to one, creating a variant we call AdaLN-One. This design preserves the original SSL feature flow at initialization, while allowing the model to gradually learn to integrate text prompts during training.

Specially, given an input audio feature $\mathbf{x}$ and a text prompt embedding $\mathbf{p}$, AdaLN fusion is defined as:

\begin{equation}
\mathbf{y} = \mathbf{x} + g(\mathbf{p}) \cdot \text{Layer}\big( (1 + \gamma(\mathbf{p})) \cdot \text{LN}(\mathbf{x}) + \beta(\mathbf{p}) \big),
\end{equation}

where $\text{LN}(\cdot)$ is Layer Normalization, $\text{Layer}(\cdot)$ denotes the attention or feedforward layer, $\gamma(\mathbf{p})$, $\beta(\mathbf{p})$, and $g(\mathbf{p})$ are modulation parameters derived from the prompt condition, initialized to zero, zero, and one, respectively.

At initialization, it reduces to the standard residual block:

\begin{equation}
\mathbf{y} = \mathbf{x} + \text{Layer}\big( \text{LN}(\mathbf{x}) \big).
\end{equation}

This allows for seamless compatibility with the pre-trained model at the start of training. As training progresses, $g(\mathbf{p})$ is learned, enabling the model to dynamically adjust the influence of text prompt modulation based on the input prompt.


Additionally, while directly fusing prompts into each transformer block is feasible, this approach becomes inefficient when handling a large number of sound classes. To overcome this limitation, we propose an encoder-decoder architecture: the encoder focuses solely on audio representation, while the decoder integrates text prompts for fusion. This design enables encoder features to be cached during inference, improving efficiency when evaluating numerous candidate sound classes.

In summary, we denote this method as FlexSED for its flexible design that leverages pre-trained models to enable efficient OV-SED.

\subsection{Negative Query Filtering}
When training FlexSED, we randomly compose positive and negative queries, as illustrated in \hyperref[fig:framework]{\cref{fig:framework}~(c)}. Positive queries correspond to labeled sound events, while negative queries refer to unlabeled events. However, we observe that \textbf{AudioSet-Strong} suffers from potential missing labels. Treating such unlabeled yet present events as negative queries can confuse the model during training, undermine prompt-event alignment, and ultimately degrade performance.

Some missing labels arise from parent-child relationships defined in the AudioSet ontology, while others fall outside this structure. For example, a clip labeled as kid-playing may omit kid-speaking, despite both events likely being present. Similarly, classes like engine and truck often co-occur but are not explicitly linked in the ontology, making them difficult to filter automatically.

While manual identification of these missing labels is feasible, it becomes impractical at scale due to the more than 400 sound classes and the vast number of possible class pairs. To address this challenge, we leverage GPT-4\footnote{\url{https://platform.openai.com/docs/models/gpt-4}} \cite{achiam2023gpt}, which excels at understanding and reasoning over complex semantic relationships, including hierarchical structures, contextual similarities, and latent associations between concepts. Specifically, we use GPT-4 to identify various intra-class relations, such as subcategories, synonymous sound events, frequent co-occurrences, and safe negatives, classes that are unrelated or only rarely co-occurring. Guided by these insights, we sample negative queries primarily from safe negatives, thereby reducing the likelihood of conflicts with positive classes and enhancing training robustness.

\section{Experimental Setups}

\subsection{Dataset}
Strongly-labeled AudioSet is a subset of AudioSet, consisting of 100k frame-level labeled real audio clips for training, with 2k reserved for validation to assist in model selection. An additional 16k frame-level labeled clips are provided for evaluation. The dataset includes 456 distinct labels, with 407 audio classes shared between the training and evaluation sets. Training is performed on all classes, while evaluation focuses exclusively on the 407 shared classes.

\subsection{Implementation Details}
We conduct experiments on audio sampled at 16~kHz. The transformer backbone is based on Dasheng-base\footnote{\url{https://huggingface.co/mispeech/dasheng-base}}, which operates on 64-dimensional log-Mel spectrograms extracted every 10~ms using a window size of 32~ms. For text encoding, we adopt the text encoder from Laion-CLAP\footnote{\url{https://huggingface.co/laion/clap-htsat-fused}}, employing simple prompts of the form ``A sound of \textit{\{class\}}'' for sound event queries, where \textit{\{class\}} refers to the sound class names defined in the AudioSet ontology.

The FlexSED transformer is initialized with the pre-trained Dasheng model, while the CLAP model remains frozen during training. FlexSED is optimized using AdamW with learning rates of $1 \times 10^{-4}$ for the decoder and $1 \times 10^{-5}$ for the encoder. A batch size of 16 audio samples and 20 prompts per audio sample is used. Among the 20 prompts, up to 10 are positive queries sampled from the labels—if more than 10 positives are available, 10 are randomly selected; if fewer, all are used. The remaining prompts are filled with negative queries randomly sampled from negative candidates. The model is trained for 10 epochs, and the checkpoint with the best validation performance is selected for final evaluation. We apply simple Mel-spectrogram augmentations \cite{nam2022filteraugment, park2019specaugment}, such as frame shift and filter augmentation, during training for the proposed FlexSED model\footnote{We leave the integration of mixup and mean-teacher into FlexSED for future work, as they are not directly compatible with the current framework.}.

\subsection{Evaluation Metrics}
Following prior studies, we adopt the PSDS1 metric \cite{bilen2020framework, ebbers2022threshold}, which captures the intersection between ground truth and detected events, emphasizing low reaction time and accurate sound event localization. Unlike earlier works that rely on coarser metrics with resolutions \cite{hershey2021benefit}, we employ this finer-grained approach for greater temporal precision. Consistent with prior research \cite{li2024self, schmid2025effective}, we omit the variance penalty, as it was originally intended for datasets with fewer and more balanced classes. Furthermore, in line with previous work, we exclude PSDS2 from this study, as it is tuned to audio tagging rather than sound event detection and is impractical\footnote{\url{https://github.com/fgnt/sed_scores_eval/issues/5}} to compute on a server with 128 GB of memory when processing over 400 classes. We apply a median filter with a window length of 7 to smooth the model predictions.

We refer to PSDS1 computed across all classes as \textbf{PSDS1\textsubscript{A}}. While this metric emphasizes accurate timing, it still penalizes false positives. In addition, we introduce a variant, \textbf{PSDS1\textsubscript{T}}, which focuses solely on target sounds. In this setting, inference is restricted to the sound classes present in each audio clip, rather than spanning all 407 classes. This reflects a practical use case in which the target sounds are already known to the user, but precise temporal localization is still required. Such a scenario highlights the model's potential for annotating weakly labeled data \cite{gemmeke2017audio} or integrating with audio-tagging models \cite{chen2022beats}.

\section{Results and Discussion}

\subsection{Comparison with Vanilla SED}

We train two vanilla SED models for comparison. First, we reproduce ATST-Frame-SED on AudioSet-Strong following the procedure outlined in the original paper \cite{li2024self}. In parallel, we train Dasheng-SED using the same vanilla multi-class classification setup and training procedure. Notably, the Dasheng-based backbone demonstrates superior overall performance and exhibits reduced sensitivity to hyperparameters compared to ATST-Frame. Based on these observations, we adopt Dasheng as the backbone for FlexSED in our experiments. Additionally, we extend a target sound separation model  for OV-SED by adding a classification layer on the pre-trained AudioSep\cite{liu2024separate}.

As shown in \cref{tab:main}, the proposed FlexSED significantly outperforms both the vanilla SED models and the target sound separation-based model. This improvement can be largely attributed to the integration of knowledge from the pre-trained CLAP and DaSheng models. CLAP enhances FlexSED’s ability to capture semantic relationships among sound classes compared to vanilla SED models, while DaSheng provides stronger temporal and acoustic modeling through its pre-trained audio transformer than AudioSep’s convolutional U-Net architecture. In addition to the performance boost, FlexSED also offers open-vocabulary capabilities for simpler usability. Furthermore, in \cref{sec:zeroshot}, we demonstrate its effectiveness in zero-shot settings and its expandability in few-shot scenarios.


\begin{table}[t]
    \centering
    \caption{Performance comparison of sound event detection methods.}
    \label{tab:main}
    \setlength{\tabcolsep}{10pt}  
    \begin{tabular}{l c | c c}
        \toprule
        \textbf{Method} & \textbf{OV-SED} & \textbf{PSDS1$_A$~$\uparrow$} & \textbf{PSDS1$_T$~$\uparrow$} \\
        \midrule
        ATST-Frame-SED & $\times$ & 0.3846 & 0.5387 \\
        Dasheng-SED                & $\times$       & 0.3986 & 0.5349 \\
        \midrule
        AudioSep-SED                & \checkmark       & 0.3260 & 0.5332 \\
        \rowcolor{lightblue} 
        \textbf{FlexSED}                    & \checkmark     & \textbf{0.4484} & \textbf{0.5863} \\
        \quad w/o Neg. Filtering   &      & 0.4366 & 0.5627 \\
        \bottomrule
    \end{tabular}
\end{table}


\subsection{Ablation Studies}
\subsubsection{Prompt Fusion Method}
As shown in \cref{tab:ablation}, we compare several fusion strategies with AdaLN-One. Specifically, AdaLN-Zero initializes the gating parameter to zero; Token Fusion \cite{bao2023all} directly inserts the prompt token into the decoder sequence; and Cross-Attention (CrossAttn) \cite{nagrani2021attention} introduces an additional cross-attention layer in the decoder blocks, using the audio embedding as the query and the prompt token as the key and value. Among these methods, AdaLN-One achieves the best performance. Cross-Attention performs slightly worse, likely due to the substantial architectural changes it introduces. AdaLN-Zero also lags behind AdaLN-One, highlighting the importance of seamless AdaLN parameter initialization for stable and effective continuous training. Token Fusion performs the worst, presumably because the pre-trained model is optimized to attend to audio embeddings and struggles to shift focus to the newly added prompt token.

\subsubsection{Encoder-Decoder Design}
We explore different configurations for splitting the original Transformer into encoder and decoder components.\footnote{Due to computational constraints and inefficiency with large number of sound classes, we do not use all blocks for decoding.}. As shown in \cref{tab:ablation}, dividing the encoder and decoder equally yields the best performance. Reducing the decoder size results in a performance drop, likely because fewer decoder layers limit the model’s ability to capture prompt-conditioned representations. 
On the other hand, increasing the decoder beyond half of the total blocks impairs performance, as too few encoder layers remain to learn prompt-independent acoustic features. Additionally, we experiment with adding two extra decoder blocks copied from the last two layers of the original model. This strategy performs worse than directly using the last two blocks as the decoder, suggesting that maintaining the original structure is important for effective continuous training. 

\subsubsection{Negative Query Filtering Strategy}
As shown in \cref{tab:main}, FlexSED without the proposed negative query filtering strategy exhibits degraded performance, underscoring the effectiveness of our method in removing potentially conflicting negative queries that could otherwise confuse training and impair model performance.

\begin{table}[t]
    \centering
    \caption{Ablation study of model design choices.}
    \label{tab:ablation}
    \resizebox{\linewidth}{!}{%
    \begin{tabular}{lcc|cc}
        \toprule
        \textbf{Fusion} & \textbf{Enc. Blocks} & \textbf{Dec. Blocks} & \textbf{PSDS1$_A$~$\uparrow$} & \textbf{PSDS1$_T$~$\uparrow$} \\
        \midrule
        Token Fusion    & First 6   & Last 6        & 0.4378 & 0.5696 \\
        CrossAttn & First 6   & Last 6        & 0.4430     & 0.5800     \\
        AdaLN-Zero      & First 6   & Last 6        & 0.4383     & 0.5717     \\
        \midrule
        AdaLN-One       & First 4      & Last 8       & 0.4374     & 0.5861     \\
        \rowcolor{lightblue} 
        \textbf{AdaLN-One}       & \textbf{First 6}   & \textbf{Last 6}        & \textbf{0.4484} &  \textbf{0.5863}     \\
        AdaLN-One       & First 8   & Last 4        & 0.4393     & 0.5795     \\
        AdaLN-One       & First 10  & Last 2        & 0.4249     & 0.5826    \\
        AdaLN-One       & First 12  & Copied Last 2  & 0.4205     & 0.5728     \\
        \bottomrule
    \end{tabular}%
    }
\end{table}


\begin{table}[t]
    \centering
    \caption{Zero-shot and few-shot performance with different numbers of excluded classes ($K_{\mathrm{out}}$). Mean retained performance ratios (R-PSDS1) are computed relative to the full-model PSDS1$^{E}_A$ and PSDS1$^{E}_T$ on excluded classes across 3 folds.}
    \label{tab:zero}
    \setlength{\tabcolsep}{14pt}
    \begin{tabular}{c c | c c}
        \toprule
        \textbf{$K_{\mathrm{out}}$} & \textbf{Setting} & \textbf{R-PSDS1$^{E}_A$ $\uparrow$} & \textbf{R-PSDS1$^{E}_T\uparrow$} \\
        \midrule
        \multirow{4}{*}{\centering 20} & zero-shot & 64.99\% & 88.27\% \\
                                      & 5-shot    & 80.23\% & 93.94\% \\
                                      & 10-shot   & 84.58\% & 95.43\% \\
                                      & 20-shot   & 87.02\% & 96.81\% \\
        \midrule
        \multirow{4}{*}{\centering 40} & zero-shot & 55.04\% & 85.74\% \\
                                      & 5-shot    & 75.74\% & 94.26\% \\
                                      & 10-shot   & 77.41\% & 96.84\% \\
                                      & 20-shot   & 82.83\% & 98.40\% \\
        \bottomrule
    \end{tabular}
\end{table}

\subsection{Zero-shot and Few-shot Performance}
\label{sec:zeroshot}

To evaluate the model’s zero-shot performance on unseen sound classes and its adaptability through fine-tuning with limited labeled data, we conduct zero-shot and few-shot experiments. Specifically, we randomly exclude $K$ sound classes from training, ensuring these classes remain unseen for both positive and negative samples. We then assess the model’s performance on these excluded classes and compare it to its performance when trained on all classes. To ensure robustness, this procedure is repeated across three random folds of unseen classes, and we report the mean retained PSDS ratios (R-PSDS1$^{E}$) relative to the full-model performance on the excluded classes. 

For few-shot learning, we fine-tune the model on the excluded classes using a small number of labeled audio clips. The decoder is updated with a learning rate of $1 \times 10^{-5}$, and the encoder with $1 \times 10^{-6}$, over 25 epochs with a batch size of 8 audio clips.

As shown in \cref{tab:zero}, the results reveal several important trends regarding the model’s generalization and adaptability. In the zero-shot setting, the model retains a significant portion of its performance on unseen classes, particularly for PSDS1$^{E}_T$, where retention rates reach around 88\% with $K_{\mathrm{out}} = 20$ and 86\% with $K_{\mathrm{out}} = 40$. In contrast, PSDS1$^{E}_A$ exhibits more pronounced degradation, dropping from around 65\% to 55\% as the number of excluded classes increases. This is expected, as PSDS$^{E}_A$ reflects both localization and classification performance, making it more sensitive to false positives and false negatives. Unseen classes can easily cause misclassification and missed detections, which jointly impact PSDS$^{E}_A$ more severely. On the other hand, PSDS$^{E}_T$ focuses primarily on timing accuracy and does not penalize false positives, making it more robust in zero-shot scenarios. 

Once few-shot fine-tuning is applied, performance improves considerably. With as few as 5 shots, PSDS1$^{E}_A$ increases to about 80\% for $K_{\mathrm{out}} = 20$ and 76\% for $K_{\mathrm{out}} = 40$. Additional labeled samples continue to boost performance, and with 20-shot fine-tuning, PSDS1$^{E}_A$ reaches roughly 87\% and 83\%, while PSDS1$^{E}_T$ surpasses 96\% and 98\%, nearly closing the gap to full training. Although larger $K_{\mathrm{out}}$ values naturally introduce greater challenges due to a broader unseen class space, fine-tuning effectively mitigates these difficulties.

Overall, these findings highlight the model’s ability to generalize reasonably well to unseen classes in zero-shot settings, especially for temporal localization, and its strong adaptability through few-shot learning to recover classification performance. This demonstrates the practicality and scalability of the proposed approach for sound event detection in dynamic and evolving environments.

\section{Conclusions}

In this work, we present FlexSED, a framework for open-vocabulary sound event detection. FlexSED effectively addresses key challenges in OV-SED by leveraging pretrained audio SSL and text models, introducing an adaptive fusion strategy for efficient prompt-audio integration, and mitigating label noise through LLM-guided pair filtering. These design choices enable FlexSED to outperform traditional classification approaches, while supporting zero-shot and few-shot capabilities for diverse real-world applications. For future work, we plan to explore strategies such as knowledge distillation \cite{schmid2025effective} to further enhance model performance, scale training to larger and more diverse audio datasets, and investigate multimodal extensions that support richer query interactions beyond text prompts.

\clearpage
\bibliographystyle{IEEEtran}
\bibliography{refs25}







\end{document}